\documentclass[aps,
prl,superscriptaddress]{revtex4}

\def\R{\hbox{\rm I \kern-5pt R}}

\def\Tr{{\rm{Tr}}}
\usepackage{amssymb,amsmath}
\begin{document}

\title{Real World Interpretations of Quantum Theory}

\author{Adrian Kent}
\affiliation{Centre for Quantum Information and Foundations, DAMTP, Centre for Mathematical
  Sciences, University of Cambridge, Wilberforce Road, Cambridge CB3 0WA,
  U.K.}
\affiliation{Perimeter Institute for Theoretical Physics,
31 Caroline Street North, Waterloo, Ontario, Canada N2L 2Y5}
\begin{abstract}
I propose a new class of interpretations, {\it real world interpretations},
of the quantum theory of closed systems. 
These interpretations postulate a preferred factorization of Hilbert space
and preferred projective measurements on one factor.
They give a mathematical characterisation of the different possible
worlds arising in an evolving closed quantum system, in which
each possible world corresponds to a (generally mixed) evolving quantum 
state.   In a realistic model, the states corresponding to different worlds should
be expected to tend towards orthogonality as different possible quasiclassical 
structures emerge or as measurement-like interactions produce different classical outcomes.  
However, as the worlds have a precise mathematical definition, real world interpretations 
need no definition of quasiclassicality, measurement, or other concepts 
whose imprecision is problematic in other interpretational approaches.   
It is natural to postulate that precisely one world is 
chosen randomly, using the natural probability distribution, 
as the world realised in Nature, and that this world's mathematical 
characterisation is a complete description of reality.  
 
\end{abstract}

\date{August 2010}
\maketitle

\section{Introduction}

The aim of this paper is to define a new class of interpretations of quantum theory. 
These interpretations are defined in their own terms and stand or fall on their own 
merits.  However, one way to motivate them, which I find
helpful and hope readers also may, is to start from the idea 
that the unitarily evolving quantum state of a closed system somehow branches 
into different possible worlds, to note some of the problems with this idea 
as traditionally understood, and then to consider a new approach
to these problems.  

The idea of branching worlds is, of course, traditionally associated with many-worlds 
interpretations of quantum theory.  According to a traditionally popular view of many-worlds
interpretations, a branch is supposed to separate into a number of sub-branches 
whenever a quantum measurement takes place, with these sub-branches corresponding
to different measurement outcomes.  More generally, different
branches are supposed to be defined by different classical events, whether 
or not these events are associated with quantum measurements as
traditionally defined.  

Some Everettians, it should be noted, argue that the universal wave 
function can and should be interpreted differently, and in particular that a 
many-worlds interpretation need not invoke the notion of branching at a
fundamental level.  This paper does not engage in that debate, since it is 
not relevant to our discussion.  The interpretation of quantum theory 
proposed here is fundamentally different from Everett's and other previously proposed 
interpretational ideas (for example \cite{everett, belleverett, modal, collapse, ch, dag, 
oxfordeverettians, bkz}), though it has some features in common with some of them.   

That said, the mathematical definitions of branching given below should also be
of interest even to a convinced believer in purely unitary quantum mechanics who sees
no fundamental role for branchings.  They can be interpreted purely phenomenologically,
as attempts to define a branching structure for a quantum theory of cosmology, 
in such a way that the structure tends to contain quasiclassical worlds, to  
the extent that the cosmological theory allows.  
Even those who are not persuaded that a quasiclassical world need be
defined by fundamental laws acknowledge that, at any rate, we experience
such a world, and that we need to be able to characterise such worlds in
order to test fundamental theory against our experience.   
To do this, we need a procedure which
starts from the theory and generates descriptions of such worlds, thus allowing us 
to derive statements of the form ``A quasiclassical world 
has property $X$ with probability $p$, according to our theory.''  
It would be very useful if this procedure could be defined by a mathematical
algorithm.   Our definitions of branching structures are proposals for
algorithms of this type.   

Returning to our fundamental goal -- offering a new class of
interpretations of quantum theory -- as already noted, we discuss several possible candidates for algorithms defining
branching structures.   These are {\it not} intended to define a
perspectival view of quantum theory incorporating several equally valid
alternative descriptions of reality.   
Our idea is rather that, if
this programme is correct, Nature has chosen one particular algorithm
-- one of our candidates -- to describe reality.     
If it turns out that no candidate gives an empirically correct
description of quasiclassical worlds within realistic models, then 
the programme will have failed (at least in its original ambition). 
If precisely one candidate does so, the programme will have 
succeeded completely. 
If more than one does so, and we cannot find any compelling principle
explaining why Nature should have preferred one over the others, nor
any empirical test that selects among them, 
then we would argue that the programme has succeeded --
it would, after all, give a realistic one world Lorentz invariant description of reality
compatible with quantum theory and with experiment and observation,
which we need and do not currently have. 
But it would leave our understanding incomplete, in the sense that
some choices are left undetermined.

\section{Interpreting a branching structure}

\subsection{Many worlds or one?}

As Bell \cite{bellonmw} and (probably many) others \cite{egak} have
noted, if the 
branching of worlds were
precisely and objectively defined, a many worlds interpretation would seem 
unnecessarily extravagant.  Given a precisely defined branching 
structure, we can just as easily define a {\it one world} interpretation 
of quantum theory.

To flesh this out, let us suppose that we indeed have an algorithm for
defining a branching structure for the universal wave function,
given a formulation of quantum theory, an initial state $\psi (0)$ at $t=0$, 
and a Hamiltonian $H$.   For the purposes of this 
illustration, we follow a traditional view of a branching structure,
taking it to be a process by which the unitarily evolving 
wave function is divided into a sum of components 
at some time $t_1$, each component then  evolves unitarily but may
further sub-divide at some later time, from which point
the sub-components then unitarily evolve until such time as
they sub-divide into sub-sub-components, and so on.  On this view,
by definition, branches remain independent: they are not allowed to
merge again after dividing. 
Suppose too that the branching structure becomes asympotically constant --
in other words that the number of branches becomes constant --  as 
$ t \rightarrow \infty$, and that we can apply the algorithm in the 
asymptotic limit to obtain a complete list of branches $B_j$.

Each branch is then uniquely identified by a final component, that is,
one which will undergo no further sub-division: the branch can be
identified throughout time by following the branching diagram 
backwards from the final states.   
Each branch $B_j$ thus 
defines an unnormalised pure quantum state $\psi_j (t)$ at each time $t$,
which can be identified from the final component by following the 
branch backwards and taking the component corresponding to the
branch at any given time. 
The branch states 
need not be, and generally will not be, orthogonal at every time
$t$: in particular, before any given sub-division, all branches
whose final components ultimately arise from that sub-division
will have the same branch state.  
However, we suppose that asymptotically the branch states  
define an orthogonal decomposition of $\psi (t)$, so that 
$$
\lim_{t \rightarrow \infty} ( \psi (t) - \sum_j \psi_j (t) ) = 0 
$$
and 
$$ \lim_{t \rightarrow \infty} ( \psi_j (t) , \psi_{j'} (t) ) = 
p_j \delta_{j j'} \, ,$$ 
where the $p_j$ are constants.  

We can then define the probability $p(B_j )$ of 
each branch $B_j$ by the Born rule.
If the relevant states also have asymptotic limits, so that we can define 
$$ 
\psi^{\infty} 
  = \lim_{t \rightarrow \infty} \exp ( - i H t / \hbar ) \psi (0) \, , 
$$
and 
$$
\psi_j^{\infty} = \lim_{t \rightarrow \infty} \, \psi_j (t) \, ,
$$
we have 
$$
( \psi_j^{\infty}  , \psi_j^{\infty} ) = p( B_j ) =  
( \psi^{\infty}  , \psi_j^{\infty} ) = p_j  \, . 
$$

More generally, so long as the relevant inner product has an asymptotic limit, 
we can define
\begin{equation}\label{ftborn}
p( B_j ) = \lim_{t \rightarrow \infty} \,  ( \psi(t) , \psi_j (t) ) \,  , 
\end{equation}
where $\psi(t) = \exp ( - i H t / \hbar ) \psi (0) $ is the unitarily 
evolved state at time $t$. 

If these assumptions --- the existence of a precisely 
defined branching structure and the existence of the relevant asymptotic limits --- held, then
a perfectly sensible interpretation of the mathematics would 
be to postulate that precisely {\it one} branch is randomly chosen, via 
the Born rule probability distribution, and that
this branch alone is realised in Nature.  

Clearly, this would be more economical than a many-worlds interpretation of
quantum theory, even if (which is hotly contested
\cite{akinbook, albertinbook}) 
we could make recover all the predictions of standard quantum theory 
from many-worlds quantum theory.   
Here, just one branch would be realised in 
Nature, rather than all.  
It would also make clear what the Born probabilities are probabilities {\it of}. 
In contrast, it is (at the very least) less immediately clear 
that many-worlds interpretations can find a sensible role for 
Born probabilities: 
whatever $p(B_j )$ might 
possibly be \footnote{See e.g. \cite{deutsch, wallaceinbook, papineauinbook}
for some proposals.}
in a many-worlds interpretation, it clearly cannot 
be the probability of reality being described by the branch $B_j$, 
since all the branches are equally real.  

That said, the aim here is not to argue that, 
given our starting assumptions,
the one-world view is demonstrably correct and the many-worlds view completely 
indefensible.   These questions are considered in detail
elsewhere \cite{akinbook, albertinbook}.  The point here is simply to observe  
that the one-world view appears natural.   
The one-world view is not actually logically necessary 
to motivate all the interpretational ideas that we set out below.
However, it will be adopted from here on, on the grounds that it is both 
pedagogically helpful and fundamentally appealing. 

Next we examine some standard intuitions about the form of the branching 
structure.   

\subsection{Do measurement theory and decoherence alone allow us to
define a satisfactory one-world interpretation?}

It might perhaps (though naively) be argued that we can
go on to sketch a satisfactory interpretation 
of quantum theory, by defining the branching structure as follows.

First, suppose that each sub-branching --- each point at which a branch
splits into two or more --- corresponds to a quantum measurement type interaction. 
Then in principle we can define the entire branching tree 
in a cosmological theory by listing all the quantum measurement-type 
interactions.  Each elementary branch can then be defined by listing 
a consistent set of possible outcomes of all the measurements encountered on
some branching path from the initial to the final time.   
If we apply the Born probability rule as above, 
then precisely one branch is picked out as corresponding to reality, and hence
precisely one set of consistent measurement outcomes is realised.  

Now, standard decoherence
arguments tell us that we can calculate the probabilities of quantum measurement outcomes 
at any point after they have been classically recorded, and that the answers will be
essentially constant after that point.  In particular, the final time 
Born probability rule (\ref{ftborn}) for branches gives essentially 
the same probability for the set of measurement outcomes
as that predicted by Copenhagen quantum mechanics.
Thus, it might be argued, we can derive the standard
Copenhagen Born rule probabilities for individual measurement events
from the final time Born rule for branches, since we have postulated that 
branchings
correspond to quantum measurements.   
Hence, the argument runs, we have outlined an interpretation
of quantum theory which treats the universe as a closed quantum
system and which explains both why we see a quasiclassical world 
and why the Copenhagen interpretation of quantum theory 
describes the outcomes of our experiments within that world.   

\subsection{Problems with naive definitions of branching}

This is too glib, though.  Invoking a ``quantum 
measurement type interaction'' as a fundamental notion reproduces
one of the unsatisfactory features of the Copenhagen interpretation.   
We may think we know roughly what we mean by the term, but we don't have a precise definition.  
And how can this explanation work without one?
Is a cosmic committee meant to survey the evolution of the 
universe and decide case by case whether, and if so precisely when,
a measurement took place?  

In any case, the notion of quantum measurement isn't sufficiently general
to produce a realistic interpretation of a quantum theory of cosmology.  
All the quasi-classical objects and structures in our universe, 
including all measuring devices, natural and artificial,
emerged from a purely quantum initial state.\footnote{And, plausibly, there is a natural sense in which 
the universe continues to become more quasi-classical: in other words,  
new quasi-classical structure
continues to emerge, and perhaps always will.}
The emergence of quasiclassical structure from a quantum
state doesn't correspond to a quantum measurement 
interaction in any standard sense of the phrase. 
Yet, without an account of this process, we can't begin to 
justify an account of the world as it now appears or 
the measurement devices and classical records it now contains. 
 
This is a deeper problem than seems to be generally recognised. 
The set of possible quasi-classical structures that could have 
emerged from the initial state of our universe is, presumably,
a continuous set, not a discrete one, at least if space-time
itself is continuous.   The peak of the Moon's centre of mass  
wave function could have been epsilonically further away 
from that of the Earth, our galaxy's relative momentum to its 
neighbours could have been slightly different, and so on.   
All these possible quasi-classical structures are 
ultimately defined by quantum states.
So we cannot hope to identify classically distinct possible final outcomes within
a cosmological theory simply by inspection; we need to find a definition that
makes sense within quantum theory.    

We could try to solve this problem mathematically      
by defining a complete set of mutually exclusive quasi-classical states, 
from which precisely one is realised, {\it if} we 
had some natural mathematical 
prescription; for instance, a natural projective decomposition 
on the subspace spanned by all the
quasi-classical states.   However, the notion of quasi-classicality
doesn't, {\it per se}, supply such a prescription.  

Another consequence of the continuous multiplicity of quasi-classical structures
is that it is hard to define precisely what we mean by a particular experiment in
the context of a continuum of possible cosmologies --- and so 
labelling the Everett branches of a cosmology by the outcome of an 
experiment is much more problematic than it first appears. 
Would an instantiation of the experiment in an alternative quasiclassical world
with the Moon's centre-of-mass wavefunction peak very slightly further 
away from the Earth's count as the same experiment or a 
different one?   What about 
an instantiation in which the experimental apparatus
is in a slightly different quasiclassical state?   Again, referring
to quasiclassicality, measurement, or similarly imprecise intuitions  
doesn't resolve the ambiguity: we need a mathematical criterion.    

\section{Characterising branches mathematically}\label{rwisec}

We are thus motivated to propose a mathematical 
characterisation of possible branches.  First, we need some assumptions.   

Suppose we are given a quantum theory of a closed system
with Hilbert space ${\cal H}$, Hamiltonian $H$ and initial 
state $\psi_I = \psi (0) $.  In non-relativistic quantum mechanics, we
can obtain the state at any later time $t$ as 
$ \psi(t) = \exp ( \frac{ - i H t }{ \hbar } ) \psi_I $.  
In relativistic quantum field theory with a suitable fixed background space-time,
for example Minkowski space, we can obtain the quantum state on any spacelike
hypersurface by the Tomonaga-Schwinger formalism, assuming
(as we do) that we have a field theory for which the 
Tomonaga-Schwinger solutions are mathematically well defined
on all hypersurfaces.    
To apply our discussion to a hypothetical quantum theory of gravity, we 
assume for now -- without worrying about the details --
that the quantum gravity theory allows something analogous, i.e. that we can evolve
the initial state to produce sequences of states of the matter and gravitational 
fields that characterise the evolving physical system.  

It might be objected that we cannot presently justify
assuming the existence of a mathematically well defined quantum
field theory that characterises the non-gravitational interactions, let alone a mathematically 
well defined quantum theory of gravity.  
Granted, theoretical physics has not produced such theories as yet
(or at least not demonstrably so). 
But our aim here is to offer a new  
way of interpreting quantum theories, not to solve all the 
current problems of theoretical physics.  We need to assume that
we have a mathematically well defined theory in order for there
to be any sense in trying to interpret it.   We also need to 
postulate some properties of the theory.  We can't be sure that these postulates
will turn out to be valid for the final quantum theory of everything, if indeed there
is one.   However, they seem reasonably
plausible, are not particularly ad hoc, and do in fact apply to 
some interesting theories.   
  
So, we adopt the following assumptions:

\vskip10pt  
{\bf Postulate 1} \qquad {\it The Hilbert space ${\cal H}$ of the system
has a natural preferred 
representation as a tensor product ${\cal H} = {\cal H}_A \otimes
{\cal H}_B$, which is defined at every time (or on every spacelike hypersurface, or
for every possible evolved state in a quantum gravity theory).  }
\vskip 10pt

{\bf Postulate 2} \qquad {\it There is a natural preferred complete set of
orthogonal projections $\{ P_i \}_{i \in I}$ 
defined on the Hilbert space ${\cal H}_B$ at every time. 
(Or on every spacelike hypersurface, etc.: to save 
repetition and fix our notation we henceforth state the postulates
for the case of non-relativistic quantum mechanics and take the 
(more fundamentally interesting) alternatives of relativistic
quantum field theory and quantum gravity as understood.)    }
\vskip 10pt
 
For quantum field theory in Minkowski space, 
examples of possible natural tensor products are given by 
taking ${\cal H}_A$ and ${\cal H}_B$ to be the subspaces
corresponding respectively to fermions and bosons, or massive and 
massless particles.  (Or vice versa -- see the discussion below.)
For a quantum gravity theory, possible examples might be for 
${\cal H}_A$ and ${\cal H}_B$ to correspond respectively 
to the gravitational and matter degrees of freedom, or perhaps
to the massless particles (including gravitons and photons) and 
the massive particles.   (Or perhaps vice versa -- again see 
the discussion below.) 

An example of a possible natural projective decomposition (defined
via a limit) is the
set of projections onto the simultaneous eigenstates of
single-particle position operators.\footnote{ That these 
eigenstates are not defined in the original Hilbert space
{\cal H} does not imply that the objects we are ultimately
interested in, the real states (\ref{realstatedef}), are not
well-defined density matrices on ${\cal H}_A$.  In fact, physical
intuition suggests they should be well-defined, since they represent
intermediate states of the ${\cal H}_A$ subsystem given a final
position measurement (of unknown outcome) on ${\cal H}_B$.}

Other examples are the sets of projections onto simultaneous eigenstates of 
momentum operators or of energy operators.
Another example, which depends on the state ${\psi (t)}$ at any
given time (or the state on any given hypersurface, etc.) is determined by the 
Schmidt decomposition
$$
{\psi (t) } = \sum_j ( p_j (t) )^{1/2}  e_j (t) \otimes  f_j (t) \, , 
$$
where for each time $t$ the sets
$ \{ \, e_j (t) \, \} $ and $ \{ \, f_j (t) \, \}$ 
are subsets of orthonormal bases of ${\cal H}_A$ and ${\cal H}_B$ 
respectively.  In this case we can take
the projectors $P_i$ at time $t$ to be projections onto the
subspaces $ \langle \, {f_j (t) } : p_j (t) = p_i (t) \, \rangle $ 
spanned by vectors $f_j (t)$ whose Schmidt coefficients are equal.  

Given the chosen projective decomposition, we can decompose
the state at any time as 
\begin{equation}\label{tdecomp}
 \psi (t)  = \sum_k  ( I \otimes P_k ) \, \psi (t) =  \sum_k \psi_k (t) \, ,
\end{equation}
where $ \psi_k (t)  = ( I \otimes P_k ) \, \psi (t) $.  
Here the range of $k$ may depend on time $t$: for instance, if the $P_k$
are Schmidt projections, the size of their set depends on the number of 
degeneracies in the Schmidt decomposition.  The notation $\sum_k$ 
should be read as implicitly allowing this, and also as allowing
integrals over continuous ranges instead of, or combined with, 
a discrete sum.  

To avoid any possible misinterpretation at this point, note that 
the branching structure we propose below will {\it not} be directly 
defined by this decomposition.  The possible branch states at time
$t$ will {\it not} be defined to be the $\psi_k (t)$, but rather
certain mixed states depending on them: see equation (\ref{realstatedef})
below.   

Suppose now that we have some hypothetical fundamental physical theory 
which stipulates that time runs only from $0$ to $T$, so that 
we have initial state $\psi_I  = \psi (0 )$ and final
state ${\psi_F } = { \psi (T) }$.   
We have the final state decomposition
\begin{equation}\label{finaldecompt}
{ \psi (T) } = \sum_l  ( I \otimes P_l ) {\psi (T)} = \sum_l \psi_l (T) \, .  
\end{equation}
We will define a set of branches with a one-to-one correspondence between
the branches $B_l$ and those states 
$ \psi_l (T) $ in equation (\ref{finaldecompt}) that are nonzero.\footnote{Here, 
to simplify the notation, we consider the case where the sum
is discrete.  If we have a continuous set of branches, then to obtain finite
quantities we need to integrate over small intervals $(l, l + \delta l)$ 
in branch parameter space, and then consider the limit as $\delta l \rightarrow 0$.}

Define
$$
q_l^T (k, t) =        | \, ( \, \psi_l (T) \, , \, \exp(-iH (T-t) / \hbar )  
{ \psi_k (t) } \, )  \, |^2 \,  
$$
and
$$
p_l^T (k, t) = \frac{ q_l^T (k, t) }{ \sum_k q_l^T (k, t) } \, . 
$$
Define
\begin{equation}\label{realstatedef}
\rho_l^T (t) = \sum_k 
{ \frac{ \Tr_{{\cal H}_B} (  \psi_k (t)  \psi_k^{\dagger} (t)  )}{ 
( \, \psi_k (t) \, , \, \psi_k (t) \, ) } }   p_l^T (k, t) \, , 
\end{equation}
where the sum is over indices $k$ for which $\psi_k (t)$ is nonzero.  
We call $\rho_l^T (t)$ the {\it real state} of branch $B_l$ at each time $t$.  
Note that $\rho_l^T (t)$ is generally a mixed quantum state and is
defined on ${\cal H}_A$ alone.  

{\bf Postulate 3} \qquad  {\it In a closed quantum system for which a 
fundamental physical theory stipulates that time
runs only from $0$ to $T$, precisely one of the branches $B_l$ is realised.
The branch $B_l$ is realised with probability
$$
p_l^T  =  \,  ( \, \psi_l (T) \, , \, \psi_l (T) \, ) \, = \, ( \, \psi_l (T) 
\, , \, \psi (T) \, ) \,  . 
$$
If $B_l$ is realised, then reality at each time $t$ in the range 
$0 \leq t \leq T$ is described by the real state $\rho_l^T (t)$.}

{\bf Asymptotic Hypothesis} \qquad {\it In the fundamental theories of closed 
quantum systems that we consider, the quantities defined above have asymptotic limits 
as $T \rightarrow \infty$.   
That is, the probabilities $ p_l^T $ tend to constants 
$p_l^{\infty}$ as $T \rightarrow \infty$ and the 
real state $\rho_l^T (t)$, for any finite $t$, tends to a constant state 
$\rho_l^{\infty} (t)$ as $T \rightarrow \infty$.}

{\bf Postulate 4} \qquad {\it In a closed quantum system, described by
a fundamental theory satisfying the asymptotic hypothesis, for which time runs 
from $0$ to $\infty$, precisely one of the branches $B_l$ is realised.
The branch $B_l$ is realised with probability given by the asymptotic
value 
$$
p_l^{\infty} = \lim_{T \rightarrow \infty} p_l^T  \, .
$$  
If $B_l$ is realised, then reality at each time $t \geq 0$ 
is described by the real state defined by the asymptotic 
limit 
$$\rho_l^{\infty} (t) =  \lim_{T \rightarrow \infty} \rho_l^T (t) \, . $$
}

\section{Discussion}\label{discuss}

\subsection{Ontology}

The real quantum state plays a crucial role in real world
interpretations, directly representing physical reality. 
Indeed, perhaps the most natural version of a real world interpretation
is to take the real quantum state as the primary physical quantity,
and to relegate the standard quantum state vector to the
status of an auxiliary
mathematical quantity, useful for calculating the possible realised
branches and their probabilities, but not necessarily itself
corresponding to anything
in reality.   In Bell's terminology, on this view, the real state 
corresponds to a beable, while the standard quantum state need not.  

While it may perhaps seem unfamiliar to suggest representing reality
directly by a time- or hypersurface-dependent density matrix, this 
is precisely what is proposed in realist many-worlds versions of
quantum theory.   The main differences here are that we have (a proposal
aimed at) a realist picture of one quasiclassical world rather than
many effectively independent such worlds, and a density matrix of
generally large rank rather than (as could be possible in many-worlds
quantum theory, given pure unitary evolution from a pure initial
cosmological state) of rank one.  The first is a positive advantage,
and it is not evident why the second should create new problems. 

That said, one could, alternatively, take some mathematical quantity
derived from the real state as representing reality (i.e. as the
beable), and simplify the picture.   We leave this interesting possibility 
for future exploration.   

The factorization of ${\cal H}$ into ${\cal H}_A \otimes {\cal H}_B$ is meant
to be encoded into the laws of nature; not 
an arbitrary choice, nor an observer-dependent construction, 
nor a split into system and environment defined for a 
particular experiment.    Moreover, the degrees of 
freedom characterised by ${\cal H}_B$ play a different
role to those characterised by ${\cal H}_A$: the ${\cal H}_B$ degrees
of freedom are auxiliary mathematical objects, which do not necessarily
have any direct correspondence to reality, whereas the ${\cal H}_A$ degrees
of freedom do directly represent reality.  

\subsection{Correspondence to reality}

Why might one hope such a picture can accurately represent
the quasiclassical world we perceive?   

\subsubsection{Role of the fundamental factorization}

Firstly, we require that the 
factorization ${\cal H} = {\cal H}_A \otimes {\cal H}_B$
allows us to characterise quasiclassical physics entirely in terms of 
degrees of freedom corresponding to ${\cal H}_A$.  

In the case of quantum field theory in Minkowski space, 
it seems fairly clear that this can be done.
If ${\cal H}_A$ and ${\cal H}_B$ are the subspaces
corresponding respectively to fermions and bosons, or massive and 
massless particles, then in either case we can represent the 
local densities of chemical species in terms of ${\cal H}_A$ degrees
of freedom.  We can then express quasiclassical variables ---
describing, for instance, the orbit of the Moon around the Earth, or 
the behaviour of apparatus and measuring devices in a two-slit experiment --
in terms of these local densities.  

Though it may seem slightly less intuitive, this is also presumably
true if we swap the roles of ${\cal H}_A$ and ${\cal H}_B$.  
In that case, the photon degrees of freedom in ${\cal H}_A$ allow
us to characterise atoms and larger aggregations of matter in 
terms of their associated electromagnetic fields, and hence to
recover a description of the quasiclassical world.

In the case of quantum gravity, the question is, as usual, less clear.
Taking ${\cal H}_A$ and ${\cal H}_B$ to correspond respectively 
to the gravitational and matter degrees of freedom allows the
spacetime to be characterised entirely by ${\cal H}_A$ degrees
of freedom.   Since any significant quasiclassical event 
involving matter eventually has a significant effect on the
gravitational field, this might perhaps suffice (though the scale of
the timelag and its ontological significance would need to be
considered carefully).  
If ${\cal H}_A$ corresponds to massless particles, including gravitons and photons,
then, as above, we can characterise matter distributions in terms of
their associated electromagnetic fields, and thus one would expect to
be able to give a direct description of both matter and metric at
classical scales.  

As we lack both a well defined quantum gravity
theory and a clear understanding of the relevant physics as the
universe tends towards its final state, we can offer nothing 
better than reasoned conjecture.   
Still, these possibilities suggest it is reasonable to hypothesise 
that an appropriate factorization might exist in quantum gravity theories.   

Given an appropriate choice of factorization, we 
can think of the degrees of freedom in ${\cal H}_A$ 
as characterising a quasiclassical system, and those 
in ${\cal H}_B$ an environment interacting with that system.   
We can also think of ourselves as, in a certain sense, contained within the system 
and distinct from the environment.  
The system and environment here are not disjoint and separated physical
entities: it is not the case that every physical degree of freedom within the region 
defined by our bodies belongs to ${\cal H}_A$.
Nonetheless, all the quasiclassical variables which we assume define
our experiences can be defined by degrees of freedom in ${\cal H}_A$ 
and, in that looser but natural sense, one may say we belong to ${\cal H}_A$.  

\subsubsection{Role of the final projective decomposition}

Part of the intuition behind defining branches via the final state
decomposition is that,
for a realistic model with realistic dynamics, an appropriate 
factorization, and an appropriate final projective decomposition, 
(i) each possible final state of the environment characterises 
a possible quasiclassical history of the system, (ii) significantly
different final states characterize significantly different quasiclassical histories,
(iii) significantly different quasiclassical histories are characterized by
significantly different final states.  In particular, 
the effects of any significant quasiclassical event are irreversibly 
recorded in the environment in such a way that the record can be 
``read'' by an appropriate abstractly specified final projective 
decomposition, such as one of those discussed above: incoming photons scatter
differently off different positions of a macroscopic pointer, leading
to distinct final states of the photon degrees of freedom, for example.  
The idea here is that unitary evolution acting on  
any two significantly distinct alternative quasiclassical outcomes --
arising naturally in cosmology, or of a quantum experiment -- should 
produce at $t = \infty$ states which are not only very nearly orthogonal
{\it but moreover are distinguished with
near certainty by an appropriate natural asymptotic projective 
decomposition on} ${\cal H}_B$.   For example, if the projective
decomposition is onto photon momentum states, the intuition is
that all the later time quantum states resulting from two 
different possible outcomes of a measurement are distinguished
by photon momenta, because the background electromagnetic field 
some time after the experiment effectively carries a  
record of the result which will persist to the end of time.\footnote{
The intuition is that this is normally true, not that it is
absolutely necessarily true of every possible experiment.
For instance, with sufficiently advanced technology, the measurement 
itself and the electromagnetic fields it generates might be made
part of a macroscopic interference experiment.   In such a case, 
a real world interpretation's description of reality might not 
necessarily imply that the measurement must have an essentially definite 
outcome in the realised branch.  We presently have no experiential data 
to suggest otherwise, of course.} 

We use the final projective decomposition to define an effective 
post-selection 
on the ensemble of possible evolutions.   One way of picturing this is 
that Nature first simulates the evolution of the 
universe using standard quantum theory, with
the initial state $\psi_I = \psi (0)$, carries out a measurement 
at $t = \infty$, defined by the final state decomposition, on
this simulated system, and then uses the result of this measurement to 
determine the branch that is realised in the actual universe. 
A less teleological and less anthropomorphic view is simply to think of the 
branches as natural structures that define a sample space for
a fundamentally probabilistic theory, which has a natural probability
distribution defined by the Born rule.   
This is a probabilistic block universe theory: it simply says that one element of the sample space is
randomly selected, and this random choice determines reality. 

\subsubsection{Role of the finite time projective decompositions} 

Suppose now that we adopt the view just mentioned, namely that our universe
is effectively post-selected by the outcome of a final projective 
measurement at $t = \infty$, defined by the asymptotic projective 
decomposition. 
One would then expect, if we aim for an interpretation that produces a 
description of reality at times between $0$ and $\infty$, that this
description should be determined by the post-selected final state
as well as the initial state. 
  
If we live in such a universe, then we have no direct knowledge of the post-selected
final measurement outcome, of course.  But, given a theory which fixes the initial state and
the Hamiltonian, we could in principle list all the possible
final measurement outcomes, and
use the standard quantum pre- and post-selection rules to infer the probabilities
of our observations (of our own experiments or of other events) conditioned 
on any given outcome of this final measurement.   Calculating the
probabilities of our observations this way, although unnecessarily complicated, would
be consistent with the usual calculations of the probabilities in
standard (non-post-selected) quantum theory: postulating a final measurement
does not affect the intervening dynamics or the overall (unconditioned) probability of
any intervening event.  

Let us imagine further that a measurement were carried out at time $t$ on ${\cal H}_B$, defined by
the projective decomposition at time $t$.  If our experience is characterised by 
${\cal H}_A$ --- i.e. if the only degrees of freedom we need to describe it  
are those of ${\cal H}_A$ --- then we learn nothing directly about the outcome of
the measurement on ${\cal H}_B$.   However, supposing for the moment that this is
the only measurement intervening between the initial and final states, we can infer
the probabilities of the various outcomes from the initial state and any possible
final state.  We can further calculate a density matrix describing 
the state of the ${\cal H}_A$ degrees of freedom at time $t$, using 
these probabilities: this density matrix depends on the final state $\psi_l$, and
hence (since we define the branch by the final state) on the branch $B_l$.  
This is precisely the calculation by which we arrive at the real state (\ref{realstatedef}).

Now we take a further step.  We calculate density matrices on ${\cal H}_A$ 
at {\it every} time $t$ using the above procedure, and we suppose that we can
give a coherent time-dependent description of evolving physical worlds
using the density matrices $\rho_l^{\infty} (t)$ corresponding to 
branch $B_l$ at each time $t$.\footnote{Recall that 
we do not suppose that the measurements at each time $t$ actually take 
place.  If they did, they would produce a non-unitary evolution 
of the quantum state, and any picture of evolving quasiclassical worlds
which emerged would be very different from that proposed here.}
Although it fits quite naturally within the quantum formalism, this postulate does not follow from standard quantum theory.  
It can ultimately only be justified empirically.   

What could constitute empirical justification?  
Roughly speaking, we would need to show that, given a realistic model, 
the proposed interpretation 
explains why we experience a persistently quasiclassical world of the 
type we do.  One good test would be whether it predicts
with high probability that the realised world will be persistently 
quasiclassical and that our world is in some sense typical 
in the set of realised worlds.\footnote{In other 
words, we see the sort of world we do because almost all
worlds are of that sort.}
A weaker, but still significant, and arguably necessary, criterion
is whether it predicts, with high probability, 
that if the realised world is quasiclassical, with large-scale structure, 
for some reasonably long time interval, then it will continue in that
state for a further long interval.\footnote{In other words, if 
the realised world contains the sort of structures
that we see for some time interval, then it will very probably persist in that 
state.  If so, we see a persistently quasiclassical world, plausibly, 
because almost all
worlds in which life could emerge in the first place are of that sort.}
  
These properties will clearly {\it not} hold for worlds 
defined by a generic choice of time-dependent projective 
decomposition on ${\cal H}_B$, nor will the asymptotic hypothesis
hold true for generic choices.  The particular choices suggested
above are chosen because it seems plausible that one or more
of them might indeed define persistently quasiclassical worlds
in realistic models, and that (for essentially the same reason)
the asymptotic hypothesis might indeed hold true in such models.   
The picture underlying this intuition is one of an ever-expanding
universe, starting from a conventional big bang and tending towards
a ``big freeze'', in which, as matter-matter interactions 
become less and less frequent, so do unpredictable quasiclassical
events and thus so should branchings of quasiclassical worlds.
While of course there are many unknowns and uncertainties, this 
picture appears consistent with our current understanding of cosmology.  

The idea then is that the split ${\cal H}_A \otimes {\cal H}_B$ and the postulated
natural projective decompositions at each time (on each hypersurface, etc)
are chosen so that the projective decomposition at any given time characterises
quasiclassical information about ${\cal H}_A$ effectively irreversibly recorded in 
degrees of freedom of ${\cal H}_B$.   
Each possible outcome of the final projective decomposition characterises
a possible branch, in which all the information necessary to characterise the
final quasiclassical state of ${\cal H}_A$ is encoded in the corresponding
outcome state in ${\cal H}_B$.    For each possible branch, we can 
define the real state at any given time $t$, a mixed state on ${\cal H}_A$ defined
by the pre- and post-selected measurement probabilities for the 
natural projective decomposition on ${\cal H}_B$ at time $t$. 
This state approximately characterises the quasiclassical information about ${\cal H}_A$
that is already, at time $t$, irreversibly encoded in ${\cal H}_B$.  

The real state of any given branch is thus defined via a procedure that is defined
by rules based on standard unitary quantum dynamics (with a specified Hamiltonian) 
and on the projection postulate --- though the real state itself neither follows 
standard unitary quantum dynamics nor standard projection postulate induced
state collapse.   

Can we, in fact, show that for one of the natural factorizations
and natural sets of projective decompositions described above, 
it is indeed the case that, to good approximation, in realistic cosmological models, 
the projective decomposition at any given time characterises
quasiclassical information about ${\cal H}_A$ irreversibly recorded in 
degrees of freedom of ${\cal H}_B$, and the asymptotic hypothesis holds?   
The question should be, at any rate, decidable, since there are rather 
few choices that have any serious claim to be regarded as natural.
The projective decomposition most studied to date in other contexts --- in particular
in investigations of modal interpretations \cite{modal}
of quantum theory --- is the 
Schmidt decomposition.  Reasons have been identified for doubting
that it generally identifies quasiclassical degrees of freedom in 
realistic models \cite{schmidtdoubts},  It seems fair to say, though,
that the question is not yet definitively resolved.  
In any case, the alternatives of projective decompositions defined
by the position and momentum eigenstates of fields in ${\cal H}_B$ look very promising.   

In short, there are good intuitive reasons to take the hypothesis seriously. 
We may, in fact, have a relatively natural solution to the measurement
problem, which essentially respects unitary quantum dynamics and Lorentz
or general covariance.   The form of the real state also means that it is 
at least not obvious that this interpretation has the sort of 
potentially worrisome ``tails problem'' that arises in modified 
dynamical theories in which suppressed components of the state vector
corresponding to unselected quasiclassical worlds acquire exponentially small
amplitudes but nonetheless retain (something like) the same mathematical
structure and information content of the selected component.   

Detailed calculations in realistic models remain to be carried out;  
they should illuminate these questions further and, one might 
reasonably hope, give persuasive evidence that a well defined real world 
interpretation can adequately describe the quasiclassical reality
we experience.

\section{Acknowledgments}  Many thanks to Jeremy Butterfield for many helpful
comments on an earlier version of the manuscript; thanks also to
Lucien Hardy, Jonathan Oppenheim, Tony Short, Rafael Sorkin, Rob
Spekkens and James Yearsley for helpful conversations. 
This research was partially supported by a Leverhulme Research
Fellowship, a grant from the John Templeton Foundation,
an fqxi mini-grant and by Perimeter Institute for Theoretical Physics.  
Research at Perimeter Institute is supported by the Government of 
Canada through Industry Canada and by the Province of Ontario 
through the Ministry of Research and Innovation.

\end{document}